\documentclass[aps,prb,twocolumn,showpacs,superscriptaddress]{revtex4}
\usepackage{graphicx}
\usepackage{hyperref}
\usepackage{natbib}
\begin{document}
\title{Electronic structure of magnetic molecules V$_{15}$: LDA+U calculations, X-ray 
emission and photoelectron spectra}
\author{D. W. Boukhvalov}
\affiliation{Forschungszentrum Juelich, D-52425 Julich, Germany}
\affiliation{Institute of Metal Physics, Russian Academy of Sciences Ural Division,
Ekaterinburg 620219, Russia}
\author{E. Z. Kurmaev}
\affiliation{Institute of Metal Physics, Russian Academy of Sciences Ural Division,
Ekaterinburg 620219, Russia}
\author{A. Moewes}
\affiliation{Department of Physics and Engineering Physics, University of Saskatchewan, 
116 Science Place Saskatoon, Saskatchewan S7N 5E2, Canada}
\author{D. A. Zatsepin}
\affiliation{Institute of Metal Physics, Russian Academy of Sciences Ural Division,
Ekaterinburg 620219, Russia}
\author{V. M. Cherkashenko}
\affiliation{Institute of Metal Physics, Russian Academy of Sciences Ural Division,
Ekaterinburg 620219, Russia}
\author{S. N. Nemnonov}
\affiliation{Institute of Metal Physics, Russian Academy of Sciences Ural Division,
Ekaterinburg 620219, Russia}
\author{L.D. Finkelstein}
\affiliation{Institute of Metal Physics, Russian Academy of Sciences Ural Division,
Ekaterinburg 620219, Russia}
\author{Yu. M. Yarmoshenko}
\affiliation{Institute of Metal Physics, Russian Academy of Sciences Ural Division,
Ekaterinburg 620219, Russia}
\author{M. Neumann}
\affiliation{Universit\"at Osnabr\"uck, Fachbereich Physik, D-49069 Osnabr\"uck,
Germany}
\author{V. V. Dobrovitski}
\affiliation{Ames Laboratory, Iowa State University, Ames IA 50011, USA}
\author{M. I. Katsnelson}
\affiliation{Institute of Metal Physics, Russian Academy of Sciences Ural Division,
Ekaterinburg 620219, Russia}
\affiliation{Uppsala University, Department of Physics,
Box 530, SE - 751 21 Uppsala, Sweden}
\author{A. I. Lichtenstein}
\affiliation{University of Nijmegen, NL-6525 ED Nijmegen, the Netherlands}
\author{B. N. Harmon}
\affiliation{Ames Laboratory, Iowa State University, Ames IA 50011, USA}
\author{P. K\"ogerler}
\affiliation{Ames Laboratory, Iowa State University, Ames IA 50011, USA}

\begin{abstract}

Electronic structure of V$_{15}$ magnetic molecules
(K$_6$[V$_{15}$As$_6$O$_{42}$(H$_2$O)]$\cdot$8H$_2$O) 
has been studied using LSDA+U band structure
calculations, and measurements of X-ray photoelectron (valence
band, core levels) and X-ray fluorescence spectra (vanadium K$\beta_5$
and L$_{2,3}$, and oxygen K$\alpha$). 
Experiments confirm that vanadium ions
are tetravalent in V$_{15}$, and their local atomic structure is close to
that of CaV$_3$O$_7$. Comparison of experimental 
data with the results of electronic structure calculations show that 
the LSDA+U method provides a description of the electronic
structure of V$_{15}$ which agrees well with experiments.

\end{abstract}

\maketitle

\section*{Introduction}

A new class of magnetic compounds, molecular
magnets, have attracted much attention due to their unusual
magnetic properties that in general are associated with 
mesoscopic-scale magnetic particles \cite{1,1a}. These materials are 
the objects of studies for 
spin relaxation in nanomagnets, quantum
tunneling of magnetization, topological quantum phase
interference, quantum coherence, etc. \cite{5,6,7}. 
In this work, we study the
polyoxovanadate K$_6$[V$_{15}$As$_6$O$_{42}$(H$_2$O)]$\cdot$8H$_2$O 
molecules (denoted below as V$_{15}$), which possess an interesting
layered
structure \cite{11,16,v15str1}. This molecule
contains fifteen
antiferromagnetically coupled vanadium ions, each having spin
$S=1/2$, see Fig.\ \ref{fig1}.
In contrast with many other molecular ferrimagnets
(such as Mn$_{12}$ or Fe$_8$), V$_{15}$ is a molecular 
antiferromagnet with small net uncompensated spin 1/2,
and it exhibits weak anisotropy.
It presents unusual features,
such as ``butterfly-like'' hysteresis loops \cite{bfly},
and, as theoretical estimates suggest \cite{2}, might exhibit
rather long decoherence time.
Previous considerations \cite{11,16,v15str1,3,4}
advocate that vanadium ions in this
compound have valency of 4+.
However, experimental studies elucidating the
electronic structure of V$_{15}$ are absent.
Such studies are important for 
theoretical considerations of this complex 
compound, and firm experimental evidence of the tetravalent
nature of vanadium ions in V$_{15}$ is crucial.

We have addressed these issues by 
investigating V$_{15}$ with
X-ray photoelectron (XPS) and X-ray emission (XES)
spectroscopies. 
These techniques allow determination of the charge (valence)
state of the ions, provide information about the total density of
states (DOS) normalized to photoionization cross-sections and
the partial DOS of atomic components in the valence band,
indicate variations in chemical bonding character, etc.
We present the complete XPS and XES study of valence states
of the V ions, and the distribution of the total and partial DOS in
the valence band of V$_{15}$. The experimental data obtained are
compared with theoretical LSDA and LSDA+U calculations, and we
show that the results obtained using LSDA+U technique 
agree well with experiments.
\section{Sample preparation and experimental details}
The XPS measurements have been carried out with a PHI 5600ci
multi-technique spectrometer using monocromatized Al K$\alpha$ radiation
(E$_{\text{exc}} = 1486.6$ eV). Estimated energy resolution is
0.35 eV, and the base pressure in the vacuum chamber during
measurements is about $5\cdot10^{-9}$ Torr. 

The XES V K$\beta_5$ spectra
($4p\to 1s$ transition) were measured using a fluorescent Johan-type
vacuum spectrometer with a position-sensitive detector \cite{9}. 
Cu K$\alpha$ X-ray radiation from the sealed X-ray tube has
been used for
excitation of the V K$\beta_5$ XES. A quartz crystal (rhombohedral
plane, second-order reflection) curved to $R = 1.8$ m was used as
an analyzer. The spectra were measured with an 
instrumental energy resolution
0.22 eV. 
The vanadium L$_{2,3}$ ($3d4s\to 2p_{1/2,3/2}$ transitions) and
oxygen K$\alpha$ ($2p\to 1s$ transition) XES have been
recorded at the Advanced
Light Source (Beamline $8.0$) employing the soft X-ray fluorescence
endstation \cite{10}. The vanadium L$_{2,3}$ and oxygen K$\alpha$ XES 
have been
measured resonantly, through the V L$_{2,3}$ and O K-edges, and 
non-resonantly (far from the threshold). The instrumental
energy resolution of the V
L and O K spectra is about 0.8 and 0.3 eV, respectively. The V $2p$
and O $1s$ X-ray absorption spectra have been measured in 
the total electron yield (TEY) mode. 

The total resolution of XES measurements is composed of
the instrumental resolution (the values given above), and
the width of the core level which depends on the lifetime of the
hole. For vanadium K$\beta_5$ XES, 
the width of the core level is about 0.79 eV, which
gives about 1.0 eV of total energy resolution. For vanadium L emission,
the core level width is about 0.8 eV, and for O K$\alpha$ XES,
the core level width is about
0.2--0.3 eV.

\begin{figure}[tbp!] 
\includegraphics[width=8cm]{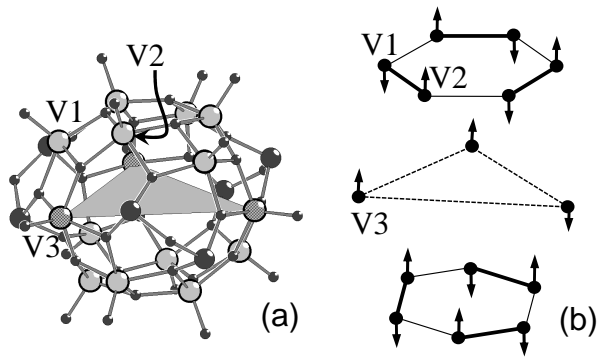}
\caption{(a) Structure of the (K$_6$[V$_{15}$As$_6$O$_{42}$(H$_2$O)]$^{6-}$
cluster, the large light-grey circles denote the vanadium ions. The 
central triangle containing the V3 ions is shaded. (b) The
schematic sketch of arrangement of the vanadium ions in V$_{15}$,
and the proposed spin ordering.}
\label{fig1} 
\end{figure}

The single crystal of 
(K$_6$[V$_{15}$As$_6$O$_{42}$(H$_2$O)]$\cdot$8H$_2$O) 
was prepared as desribed in Ref.\ \onlinecite{11}. It has a trigonal
symmetry (space group $R3c$), as shown in Fig.\ \ref{fig1}a. The overall
structure consists of three sets of non-equivalent
vanadium atoms V1, V2, and V3. V1 and V2 belong to two
nonplanar hexagons separated by a triangle of V3 centers forming
the ``layer structure'' (Fig.\ \ref{fig1}b).

\section{Discussion of experimental results}
The X-ray emission valence spectra originate from the
electron transitions between the valence band and the core
hole. The wave functions of the core states are
strongly localized, and the
angular momentum symmetry of the core electrons is well defined.
Thus,
according to the dipole selection rules, these spectra reflect
the site-projected, and symmetry restricted partial densities of
states (DOS). However, it is rather difficult to extract 
information about the occupied V $3d$ states from non-resonant V 
L$_{2,3}$ XES because V L$_3$ ($3d4s\to 2p_{3/2}$) and V 
L$_2$ ($3d4s\to 2p_{1/2}$) transitions are
strongly overlapped due to the small spin-orbital splitting of 
V $2p$ states (7.7
eV). To overcome this difficulty, we have used resonant excitation
of V L$_{2,3}$ XES, where the energy of incoming photons was tuned near
V $2p$-thresholds, and in this way we could selectively excite V 
L$_3$ XES. 

\begin{figure}[tbp!] 
\includegraphics[width=8cm]{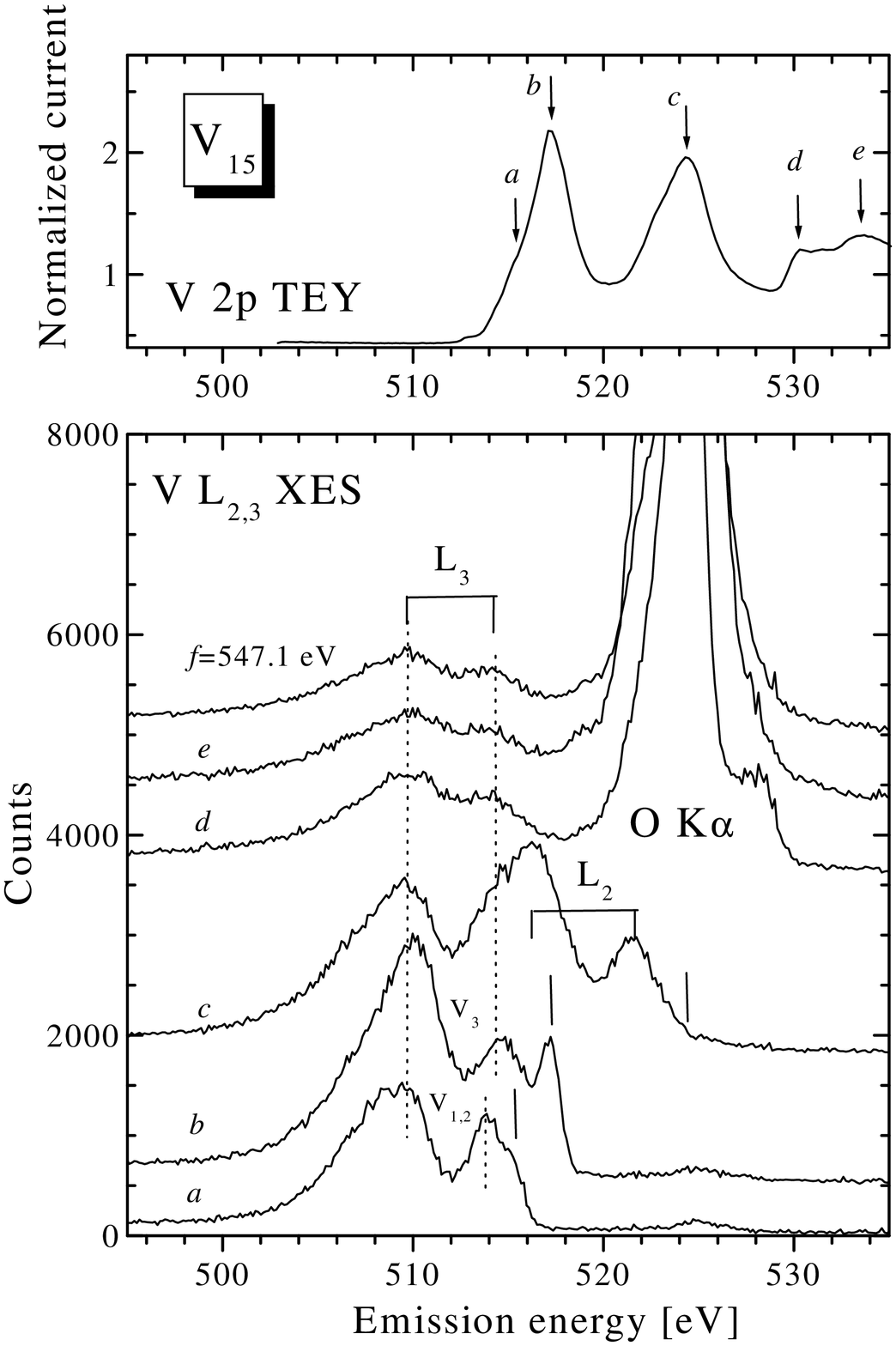}
\caption{Excitation energy dependence of V L$_{2,3}$ XES of V$_{15}$}
\label{fig2} 
\end{figure}

Fig.\ \ref{fig2} shows the results of measurements of 
V L$_{2,3}$ XES resonantly excited near the V $2p$-thresholds. 
The excitation
energies were selected in accordance with the features ($a$--$f$) on V $2p$ 
TEY (upper panel of Fig.\ \ref{fig2}), and indicated by vertical lines on V
L-emission spectra (lower panel of Fig.\ \ref{fig2}). These energies
exactly correspond to the energy of the elastic peaks which probe the
unoccupied states, and the resulting intensities follow the
absorption cross section. As one can see, the selectively excited V 
L$_3$
XES (curves $a$ and $b$, Fig.\ \ref{fig2}) reveal two peaks at 510.0 eV,
and at 513.8--514.4 eV, which reproduce the structure of the undistorted
distribution of V $3d$ states in the valence band, and can be
related to O $2p$ and V $3d$ bands, respectively \cite{13}. The V 
$3d$-like
peaks have different emission energies (513.8 and 514.4 eV), which
can be attributed to the contributions of non-equivalent vanadium
atoms in the crystal structure of 
K$_6$[V$_{15}$As$_6$O$_{42}$(H$_2$O)]$\cdot$8H$_2$O
(Fig.\ \ref{fig1}). The ratio of V $3d$ and O $2p$ (V $3d$)-like peaks is higher
for the curve $a$ than for the curve $b$, which allows us to
associate the peak at 513.8 eV with the contributions of V1 and V2 atoms
from hexagons (we denote this contribution as 1-2), and the peak at
514.4 eV --- with the contribution of V3 atoms (we denote this
contribution as 3). At higher excitation energies, the peaks at
513.8--514.4 eV are strongly overlapped with the main peak of V 
L$_2$ XES (curves $c$--$f$, Fig.\ \ref{fig2}),
which makes it very
difficult to establish their location and width. 

\begin{figure}[tbp!] 
\includegraphics[width=8cm]{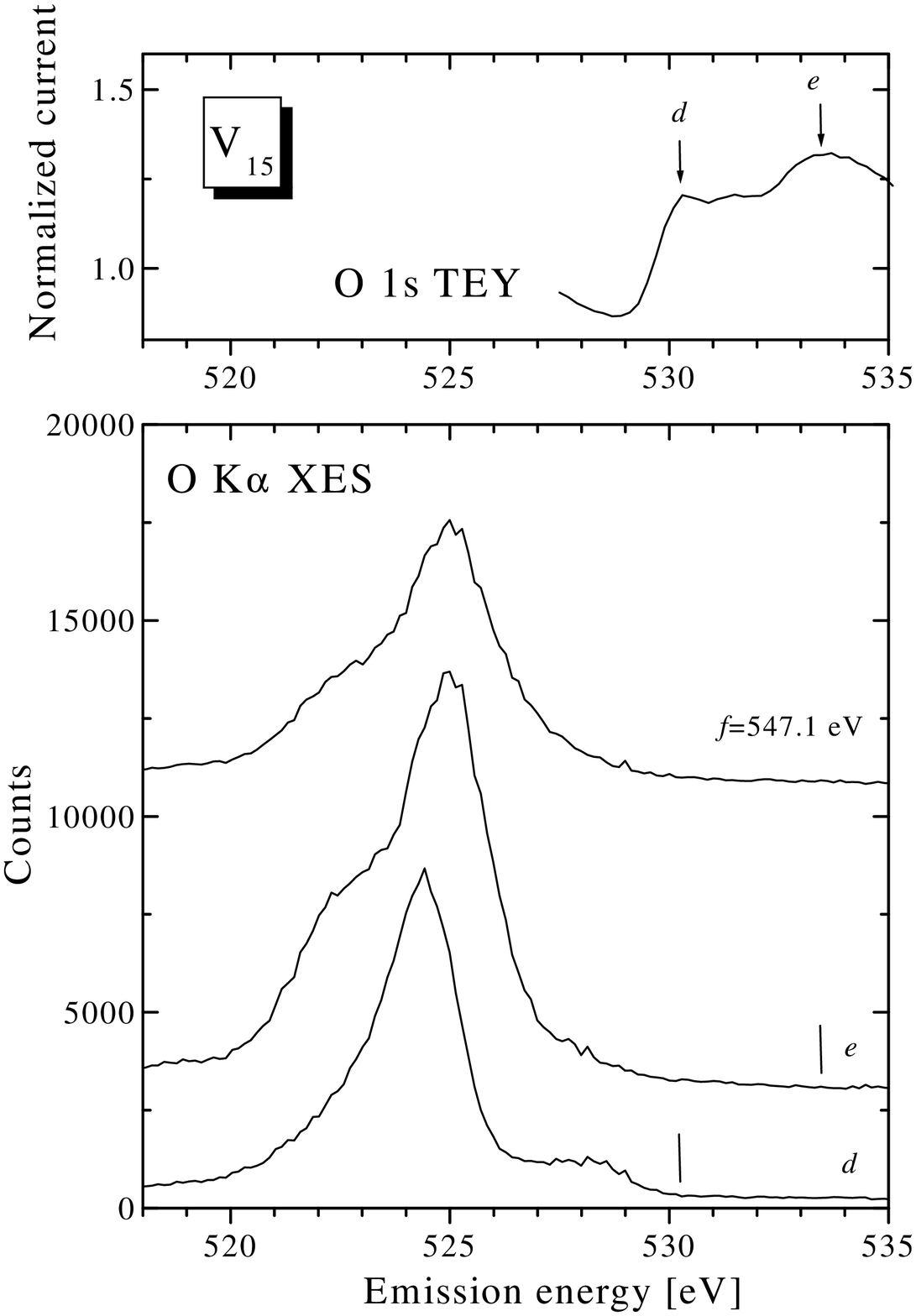}
\caption{Excitation energy dependence of O Ka XES of V$_{15}$ }
\label{fig3} 
\end{figure}

The
resonantly excited O K$\alpha$ XES of V$_{15}$ (Fig.\ \ref{fig3}) 
reveal significant
changes in the fine structure, depending on excitation energy.
These changes can be explained by contributions of different
oxygens belonging to the polyoxovanadate part and the water of hydration in
the V$_{15}$ structure, which are selectively excited by tuning the
energy of incoming photons. Using the X-ray
fluorescence measurements of liquid water \cite{12}, we can attribute
the feature $e$ of O $1s$ TEY (upper panel, Fig. 3) and corresponding
O K$\alpha$ XES (curve $e$, lower panel, Fig.\ \ref{fig3}) to 
the oxygens belonging to
the water of hydration. We believe that the curve $d$ of O K$\alpha$ 
XES (lower
panel, Fig.\ \ref{fig3}) corresponds to the contribution of
oxygen atoms from the polyoxovanadate part of V$_{15}$. 

\begin{figure}[tbp!] 
\includegraphics[width=8cm]{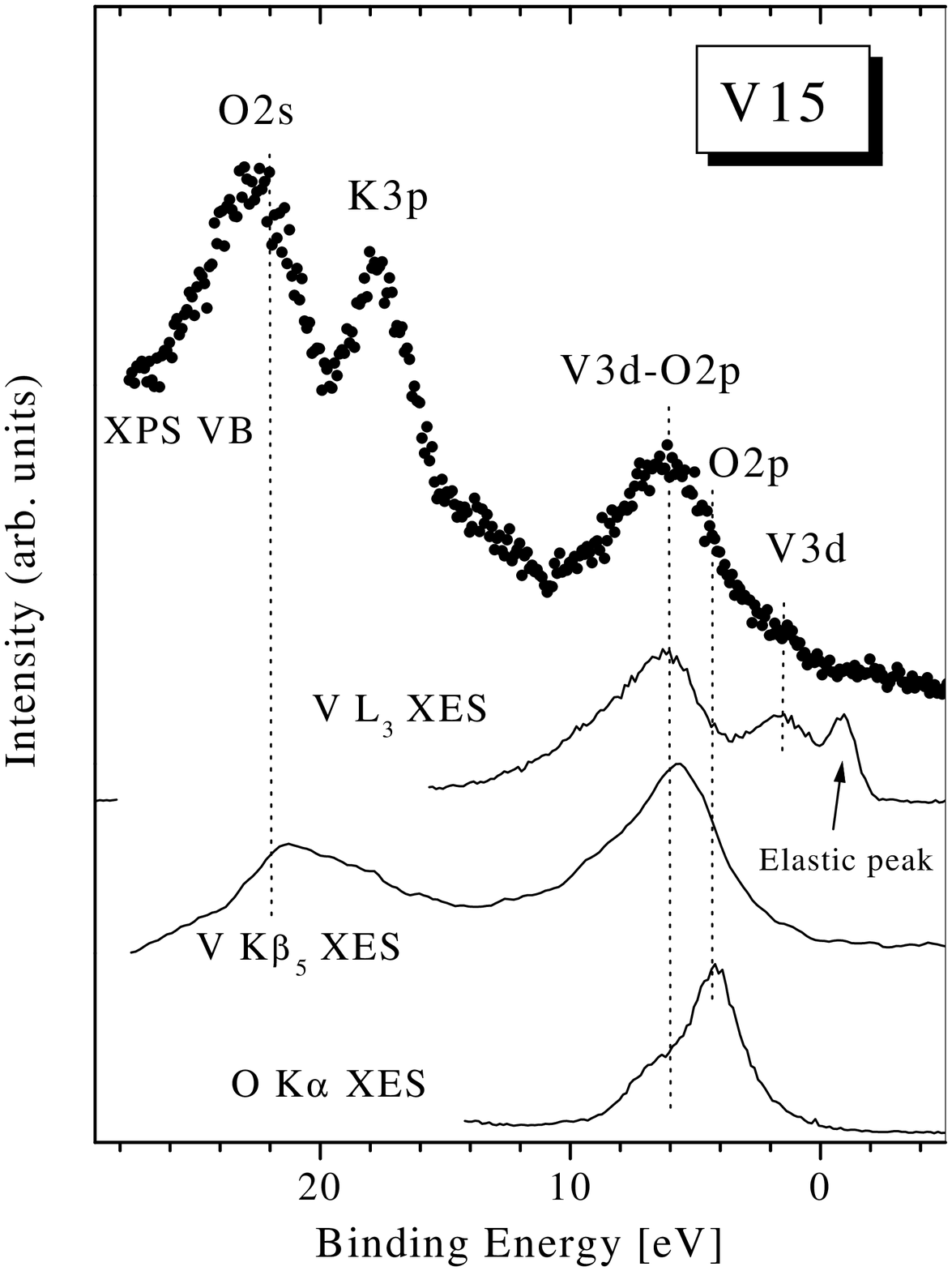}
\caption{The comparison of XPS VB and XES of constituents of 
V$_{15}$ in the binding energy scale}
\label{fig4} 
\end{figure}

In Fig.\ \ref{fig4}, we compare the  
XPS VB and X-ray emission spectra for
V K$\beta_5$, V L$_3$, and O K$\alpha$ XES. To
convert the X-ray emission spectra to the binding energy scale, we
have used XPS binding energies of core levels (V 2$p_{3/2}$, O $1s$),
emission energy of V K$\alpha_1$ ($2p_{3/2}\to 1s$ transition), 
V L$_3$ ($3d\to 2p_{3/2}$
transition) and O K$\alpha$ ($2p\to 1s$ transition) 
measured for V$_{15}$. Such a
comparison provides a direct interpretation of XPS VB (which
probes total DOS), because the XES of the constituents probe partial
DOS's due to the dipolar selection rule. As one can see, 
at the bottom of
the valence band, the O $2s$ ($\sim 22$ eV) and K $3p$ states 
($\sim 18$ eV) are
located, as revealed in the XPS VB. V K$\beta_5$ XES also
consist of an energy band (K$\beta''$) around 21 eV, 
because of the hybridization between V $4p$ and O $2s$ states.
Such hybridization is typical for all vanadium oxides \cite{13}.
According to Fig.\ \ref{fig4}, in the middle of the valence band 
V $3d$ (V $4p$)
and O $2p$ states are concentrated, and strong mixing is
present. Our
spectra demonstrate that at the top of the valence band, the 
V $3d$ states prevail. 

\begin{figure}[tbp!] 
\includegraphics[width=8cm]{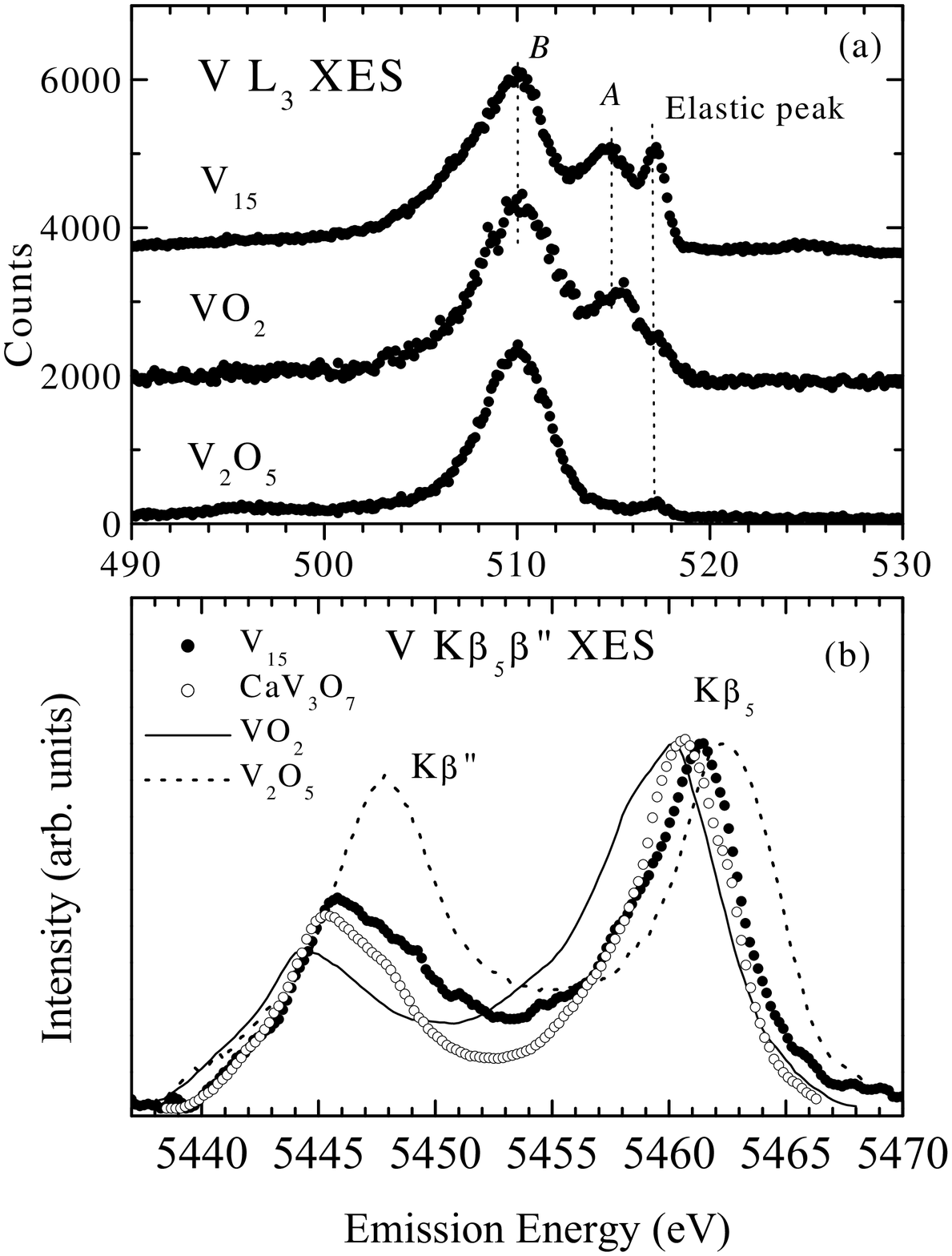}
\caption{The comparison of V L$_3$ (a) and V K$\beta_5$ (b) of 
V$_{15}$ with spectra of reference samples}
\label{fig5} 
\end{figure}

To analyze the oxidation state of V ions and their local
atomic structure, we have compared V L$_3$ and 
V K$\beta_5$ XES of V$_{15}$
with the corresponding spectra of the reference samples 
VO$_2$ (where V ions have valency 4+), V$_2$O$_5$ (where
V ions have valency 5+) and CaV$_3$O$_7$ (where V ions
have valency 4+) (Figs.\ \ref{fig5}a,b). 
Fig.\ \ref{fig5} shows that V K$\beta_5$ and V L$_3$ XES of
V$_{15}$ are closer to those of VO$_2$ than to V$_2$O$_5$,
which suggests
that vanadium ions are tetravalent in V$_{15}$. We need to
point out that the relative intensity of elastic peak with
respect to the main peak of the L spectrum of V$_{15}$ is much higher
than that of VO$_2$ (Fig.\ \ref{fig5}a),
although the vanadium ion has the same
oxidation state (4+) in both compounds. The elastic peak
corresponds to the transitions from vacant $d$-states occupied by
a photoelectron in the intermediate state of the absorption-emission
process. We can assume that its high intensity indicates that 
the $d$-electrons in V$_{15}$ are more
localized as a result of larger
V--V distances. Indeed, the average V--V distance in VO$_2$ is about
2.62 \AA, whereas in V$_{15}$ the distance V1--V2 is 2.86 \AA and
3.063 \AA \cite{16}. 
The V3--V3 distances (6.926 \AA) in V$_{15}$ are 
even larger. 

It is 
known that the spectral parameters of V K$\beta_5$ XES (the energy
position, the K$\beta''/{\text{K}}\beta_5$ intensity ratio) 
are very sensitive not only
to the oxidation state of V ions in compounds, but also to their
nearest neighbors \cite{13}. Therefore, we have
chosen CaV$_3$O$_7$ (V$^{4+}$ oxidation state) 
and V$_2$O$_5$ (V$^{5+}$ oxidation state) as the reference samples:
in these compounds, as well as in V$_{15}$, vanadium atoms are situated
inside a distorted pyramid made by five oxygen
atoms \cite{14,15}. Comparison of the V K$\beta_5$ spectra of 
V$_{15}$, VO$_2$, V$_2$O$_5$, and CaV$_3$O$_7$ (Fig.\ \ref{fig5}b) 
shows that the
energy position, the K$\beta''/{\text{K}}\beta_5$ intensity ratio,
and the spectrum shape of 
V$_{15}$ are practically identical to the spectrum of CaV$_3$O$_7$, 
but significantly differ from the V$_2$O$_5$
spectrum. This gives 
experimental confirmation that vanadium ions in V$_{15}$ 
have the oxidation state (4+) and the configuration of 
neighboring atoms is similar to that of CaV$_3$O$_7$.
\section{Discussion of the results of electronic structure
calculations}
To obtain deeper understanding of 
the electronic structure of V$_{15}$,
and to compare the experimental data with theoretical 
predictions, we have performed a series of LSDA+U calculations
\cite{ldau}. This method is known to provide a good
theoretical description for most metal-oxide 
crystalline systems \cite{MnO,MnO1}, since in most metal-oxide 
crystals the account of the on-site Coulomb repulsion
(characterized by the value of the parameter $U$)
is crucial for correct description of their properties.
Moreover, this method has been successfully 
applied \cite{mn12ldau} for the
molecular magnet Mn$_{12}$, and account of the
on-site Coulomb repulsion yields the
correct value for the gap in the electronic spectrum.
Therefore, we expect that the LSDA+U approach
should also be successful in describing the electronic structure of
V$_{15}$. As shown below, this is indeed the case.
Details of application of the LSDA+U technique to electronic structure 
calculations of molecular magnets are described in
Ref.\ \onlinecite{mn12ldau}.

To make the calculations feasible and reasonably precise,
we have followed standard practice \cite{mn12ldau,ellis}, 
excluding
from consideration the molecules of water of hydration, but
retaining completely the polyoxovanadate part
of the V$_{15}$ molecule. The positions of constituent ions
have been obtained from the X-ray data.
The calculations presented below have been made for
$J=0.8$ eV and for $U=4$ eV. The calculations for 
other values of $U$, from 3.8 eV to 5.4 eV, do not
significantly change the DOS, but the distance between
the bands increases with $U$.

\begin{figure}[tbp!] 
\includegraphics[width=8cm]{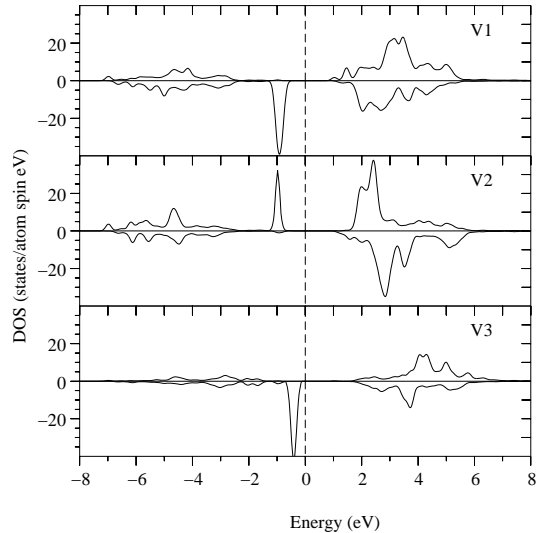}
\caption{DOS of $d$-electrons of inequivalent V1, V2, and V3 ions}
\label{figVd} 
\end{figure}

We note that our calculations with $U=J=0$ (which 
coincide with the ASA-LMTO LSDA approach) do not
give correct results, exhibiting
qualitatively erroneous non-zero DOS at the Fermi level.
As reported in Refs.\ \onlinecite{4,4a}, the use of 
LSDA approach with Gaussian basis and GGA functional,
along with theoretical optimization of the structure of
V$_{15}$ molecule, gives a gap between the
occupied and non-occupied states, with zero DOS at
the Fermi energy. As can be concluded from Ref.\ \onlinecite{4a},
this approach does not agree with some details 
present in XES spectra, e.g.\ the O $2p$ and V $3d$ bands are 
separated by larger interval than obtained experimentally, 
and than given by our LSDA+U calculations with $U=4$ eV.  
The situation is not yet clear, further work is needed
to clarify the details of the electronic structure of 
V$_{15}$, therefore we do not discuss this issue further
here.

\begin{figure}[tbp!] 
\includegraphics[width=8cm]{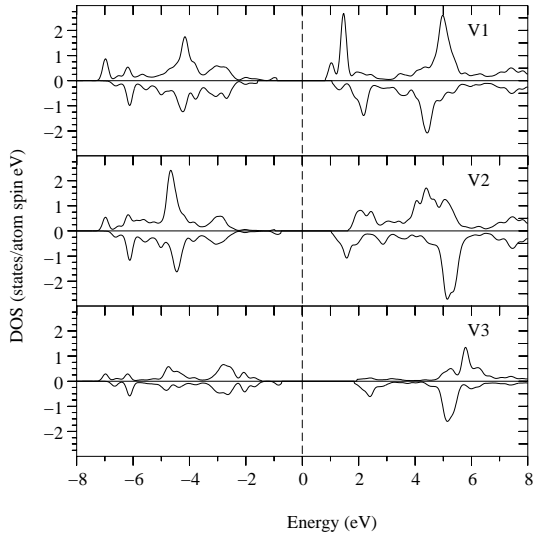}
\caption{DOS of $p$-electrons of inequivalent V1, V2, and V3 ions}
\label{figVp} 
\end{figure}

\begin{figure}[tbp!] 
\includegraphics[width=8cm]{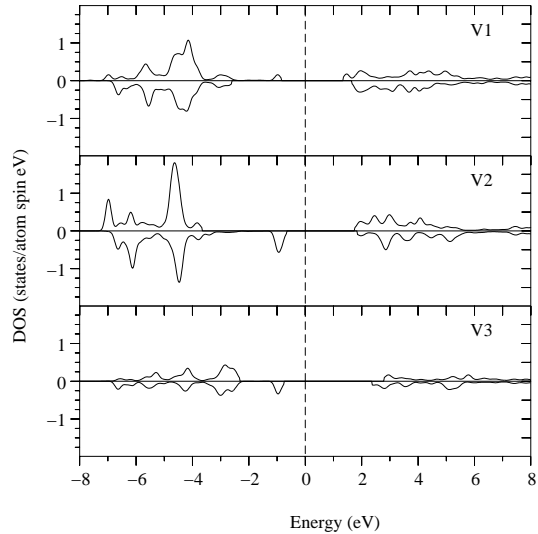}
\caption{DOS of $s$-electrons of inequivalent V1, V2, and V3 ions}
\label{figVs} 
\end{figure}

\begin{figure}[tbp!] 
\includegraphics[width=8cm]{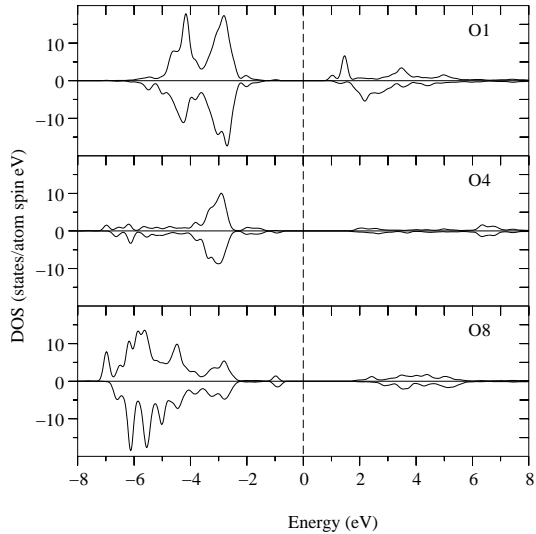}
\caption{Total DOS of $p$-electrons of oxygens belonging to 
the polyoxovanadate part of V$_{15}$ molecule}
\label{figOp} 
\end{figure}

\begin{figure}[tbp!] 
\includegraphics[width=8cm]{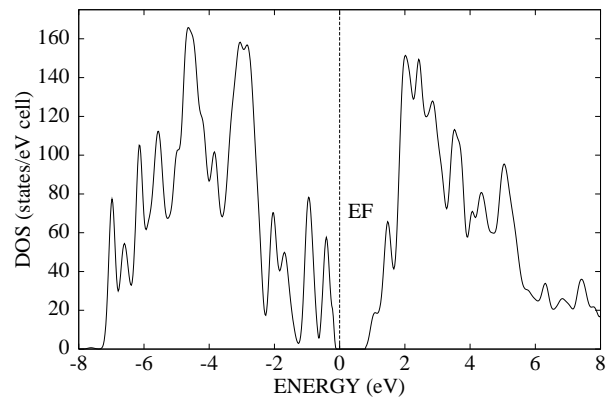}
\caption{Total electronic DOS of the polyoxovanadate part of 
V$_{15}$ molecule}
\label{figtot} 
\end{figure}

The calculated DOS of $d$, $p$, and $s$ electrons of inequivalent 
V1, V2, and V3 ions
are presented in Figs.\ \ref{figVd},\ref{figVp},\ref{figVs}, and the DOS of
oxygen ions belonging to the polyoxovanadate part of the 
V$_{15}$ molecule are shown in Fig.\ \ref{figOp}. 
The total DOS of electrons in the V$_{15}$ molecule is presented
in Fig.\ \ref{figtot}. Until now, detailed quantitative
experimental information on
the electronic structure of V$_{15}$ has been
lacking, but the results of our calculations agree
with available qualitative experimental facts.
E.g., a finite gap ($\Delta E\sim 1$ eV) in 
the total DOS is correctly reproduced by the LSDA+U calculations.
In this work, using the results of XES and XPS investigations described
above, we can make 
quantitative comparison between theory and
experiment.

The $d$ electrons of the vanadium ions determine the magnetic
behavior of V$_{15}$. Previous magnetic measurements
\cite{11,16,v15str1},
and the results of XES/XPS studies presented above,
confirm that V ions are tetravalent, with the 
total spin 1/2 per ion. 
Moreover, the measurements
of dc spin susceptibility and EPR data suggest that the 
intra-molecular exchange interactions between V1 and V2 (belonging
to the upper and lower hexagons) are strong, while
the exchange with V3 ions is much smaller.

These facts agree well with our theoretical results.
The calculated $3d$ DOS of all vanadium
ions (see Fig.\ \ref{figVd}) demonstrate
two pronounced features: the sharp
peaks located at about $-1$ eV from the Fermi level
for V1 and V2 (for V3, about $-0.5$ eV),
and the broad bands extended between
$-2$ eV and $-7$ eV. The sharp peaks correspond to localized
V $d$-electrons responsible for formation of well-defined
local spin $S=1/2$ of vanadium ions. Our calculations,
indeed, give the values of magnetic moments very close to
1$\mu_B$, namely:
\begin{eqnarray}
\mu &=& -0.94\mu_B\qquad\text{ for V1},\\ \nonumber
\mu &=& 0.91\mu_B\qquad\text{ for V2},\\ \nonumber
\mu &=& -1.0\mu_B\qquad\text{ for V3}.
\end{eqnarray}
The broad bands in the spectrum of V $d$ electrons 
clearly demonstrate the signatures of 
hybridization between the V $d$ and $s$ states, on one hand, 
and the O $p$ states, on the other. 
The broad structure of
O $p$ DOS is reproduced in V1 and V2 $d$ and $s$ DOS,
and, somewhat weaker, in V3 $d$ and $s$ DOS.
This is in agreement with 
the fact that magnetic superexchange interactions between
V1 and V2 (located in upper and lower hexagons) are very
strong ($\sim 800$ K, according to Ref.\ \onlinecite{16}),
and involve strong hybridization between V $3d$ and
O $2p$ orbitals, while the interactions of V3 (located
in the central triangle) are much weaker, implying weaker
hybridization. Similar hybridization signatures
between V $s$ orbitals
and O $p$ orbitals (see Figs.\ \ref{figVs},\ref{figOp})
correspond to chemical bonding of V with surrounding oxygens.

The calculated DOS are in agreement with the results of XES
measurements. The two features of vanadium DOS's, the sharp
peaks and the broad bands, correspond to the two wide peaks,
at about $-1.5$ eV and $-6$ eV clearly seen
in the vanadium L$_3$ XES spectrum in Figs.\ \ref{fig2},\ref{fig4}.
In agreement with experimental data, the sharp 
``magnetic'' peak in V3 is closer to the Fermi level, 
than the ``magnetic'' peaks of V1 and V2 (Fig.\ \ref{fig2},\ref{figVd}).
The difference between the peak widths and intensities
in the calculated DOS and in the measured XES spectra can
be attributed to the difference in matrix elements 
corresponding to different V $d$ states. It is known
that such a difference can be very large \cite{nemoshk},
and it can lead to significant differences between
the ``bare'' peak widths/heights in the DOS and 
the widths/intensities of corresponding peaks as observed in XES
spectra. Therefore, following the common practice,
we have restricted our discussion to the peak positions,
rather than widths and heights.
Furthermore, the structure of calculated V $p$ states DOS is 
in good agreement with the XES spectra, Figs.\ 
\ref{figVp},\ref{fig4}. The broad band
extending between $-2$ eV and $-8$ eV is revealed in the V K$\beta_5$
XES spectra as a wide peak located in the same energy interval.
Similarly, the calculated DOS for O $p$ states (a broad band
from $-2$ eV to $-8$ eV) exhibits
itself as a broad feature in the O K$\alpha$ XES spectra in the same
energy range, see Figs.\ \ref{fig3},\ref{fig4}.

The X-ray emission is a selective tool which probes
only the states allowed by dipolar selection rules, but
XPS probes the total density of states. The calculated
total DOS (Fig.\ \ref{figtot}) exhibits a number of
peaks and dips. 
But the XPS spectra, apparently, lacks
sufficient energy resolution to see this fine structure, since the
resolution of XPS measurements on dielectric
crystals, such as V$_{15}$, is considerably reduced due to the
charging effect. Therefore,
the total DOS in the energy range of interest
(between the Fermi level and $-8$ eV) is seen as a single
hump of the shape similar to the smeared total DOS as
obtained from LSDA+U calculations.
This is in qualitative
correspondence with the fact that the ``magnetic'' peaks in
the V $3d$ states constitute only a small fraction of the total
states in V$_{15}$. Along with the insufficient resolution,
other reasons might reduce the intensity of ``magnetic'' peak,
e.g.\ contamination of the surface (XPS, in contrast with
XES, is surface-sensitive, and the peak seen in the 
bulk XES spectra, can be suppressed if the surface is 
spoiled), or the final state effects.
\section{Summary}
In this work, we have performed an extensive experimental
and theoretical investigation of the electronic structure of 
V$_{15}$ magnetic molecules. Using XPS and XES spectroscopies,
we have confirmed that vanadium ions
are tetravalent, and their local atomic structure is close to
that of CaV$_3$O$_7$. We have managed to separate, identify,
and study in detail
the contributions from inequivalent 
vanadium ions V1,2 (belonging to the upper and lower hexagons),
and V3 (located in the triangle sandwiched between the hexagons)
to the XES spectra.
For theoretical studies, we have
employed the LSDA+U approach with $U=4$ eV, which results in
agreement with the experimental data. 

We found good agreement between the experimental data and 
the results of calculations. Our calculations also confirm
the oxidation state $4+$ of vanadium ions in V$_{15}$. The
main features of the calculated DOS correspond to
the peaks seen in XES and XPS spectra for V and O ions.
High intensity of the elastic peak of V L$_3$ XES indicates 
the existence of
localized states of V1, which are due to the peculiar crystal 
structure of 
V$_{15}$, where the V--V distances are much larger than 
the V--X ones. The
calculated magnetic moments of V1, V2, and V3 are very close to
1$\mu_B$, i.e.\ to the moment of a free V$^{4+}$
ion, in contrast with other compounds of tetravalent vanadium. 
\begin{acknowledgments}
Funding by the Russian Foundation for Basic Research (Projects
00-15-96575 and 02-02-16674), NATO Collaborative Linkage Grant
(PST.CLG.978044), and the Natural Sciences and Engineering
Research Council of Canada (NSERC) is gratefully acknowledged.
The work is partially supported by the Netherlands Organization for
Scientific Research, NWO project 047-008-16. This work was partially carried
out at the Ames Laboratory, which is operated for the U.\ S.\ Department of
Energy by Iowa State University under Contract No.\ W-7405-82 and was
supported by the Director of the Office of Science, Office of Basic Energy
Research of the U.\ S.\ Department of Energy.
\end{acknowledgments}

\begin{thebibliography}{99}
\bibitem{1} O. Kahn, {\it Molecular Magnetism\/} (VCH, New York, 1993)
\bibitem{1a} D. Gattesehi, A. Caneschi, L. Pardi, and R. Sessoli, Science, 
  {\bf 265}, 1054 (1994)
\bibitem{5} {\it Quantum Tunnelling of Magnetization\/}, 
 (Eds. L. Gunther and B. Barbara), NATO ASI Ser. E, Vol.301 (Kluwer, Dordrecht, 1995)
\bibitem{6} J. R. Friedman, M. P. Sarachik, J. Tejada, and R. Ziolo, 
 Phys. Rev. Lett. {\bf 76}, 3830 (1996)
\bibitem{7} L. Thomas, F. Lionti, R. Ballou, D. Gattesehi, R. Sessoli, 
 and B. Barbara, Nature, {\bf 383}, 145 (1996)
\bibitem{11} A. M\"uller and J. D\"oring, J. Angew. Chem. Int. Ed. Engl. 
 {\bf 27}, 1721 (1988)
\bibitem{16} D. Gatteschi, L. Pardi, A.L. Barra, A. M\"uller, and 
 J. D\"oring, Nature {\bf 354}, 465 (1991). 
\bibitem{v15str1} A.-L. Barra, D. Gatteschi, L. Pardi, A. M\"uller,
  and J. D\"oring, J. Am. Chem. Soc. {\bf 114}, 8509 (1992)
\bibitem{bfly} I. Chiorescu, W. Wernsdorfer, A. M\"uller, H. B\"ogge, 
  and B. Barbara, Phys. Rev. Lett. {\bf 84}, 3454 (2000)
\bibitem{2} V. V. Dobrovitski, M. I. Katsnelson, and B. N. Harmon, 
Phys. Rev. Lett. {\bf 84}, 3458 (2000)
\bibitem{3} C. Raghu, I. Rudra, D. Sen, and S. Ramasesha, Phys. Rev. B {\bf 64},
 064419 (2001)
\bibitem{4} J. Kortus, C. S. Hellberg, and M. R. Pederson, Phys. Rev. Lett.
 {\bf 86}, 3400 (2001)
\bibitem{9} V. E. Dolgih, V. M. Cherkashenko, E. Z. Kurmaev, D. A. Goganov, 
 E. K. Ovchinnikov, and Yu. M. Yarmoshenko, Nucl. Instrum. Methods. Phys. Res. A
 {\bf 224}, 117 (1984)
\bibitem{10} J. J. Jia, T. A. Callcott, J. Yurkas, A. W. Ellis, F. J. Himpsel , 
 M. G. Samant, J. St\"ohr, D. L. Ederer, J. A. Carlisle, E. A. Hudson, 
 L. J. Terminello, D. K. Shuh, and R. C. C. Perera, Rev. Sci. Instrum. 
 {\bf 66}, 1394 (1995)
\bibitem{12} A. Augustsson, J.-H. Guo and J. Nordgren, MAX-lab Activity Report, 
 p. 152, 2000, Lund, Sweden
\bibitem{13} E.Z. Kurmaev, V. M. Cherkashenko, and L.D. Finkestein, 
 {\it X-ray emission spectra of solids\/} (Moscow, Nauka, 1988) 
\bibitem{14} H. G. Bachman, F. R. Ahmed and W. H. Barnes, Z. Krist. 
 {\bf 115}, 110 (1961)
\bibitem{15} J.-C. Bouloux and J. Galy, Acta Cryst. B {\bf 29}, 269 (1973)
\bibitem{ldau}  V. I. Anisimov, F. Aryasetiawan, and A. I. Lichtenstein, J.
Phys.: Condens. Matter {\bf 9,} 767 (1997)
\bibitem{MnO} I. V. Solovyev and K. Terakura, Phys. Rev. B {\bf 58,} 15496
(1998)
\bibitem{MnO1} I. A. Nekrasov, M. A. Korotin, and V. I. Anisimov,
cond-mat/0009107
\bibitem{mn12ldau} D. W. Boukhvalov, A. I. Lichtenstein, V. V. Dobrovitski,
  M. I. Katsnelson, B. N. Harmon, V. V. Mazurenko, and
  V. I. Anisimov, Phys. Rev. B {\bf 65}, 184435 (2002)
\bibitem{ellis}  Z. Zeng, D. Guenzburger, and D. E. Ellis, Phys. Rev. B {\bf %
 59}, 6927 (1999)
\bibitem{4a} J. Kortus, M. R. Pederson, C. S. Hellberg, and
  S. N. Khanna, Eur. Phys. J. D {\bf 16}, 177 (2001).
\bibitem{nemoshk} V. V. Nemoshkalenko and V. G. Aleshin,
 {\it Electron spectroscopy of crystals\/} (New York, Plenum Press, 1979).
\end{thebibliography}
\end{document}